\documentclass[journal=jacsat,manuscript=article]{achemso}
\setkeys{acs}{articletitle = true}
\setkeys{acs}{maxauthors = true}

\usepackage{chemformula} 
\usepackage[T1]{fontenc} 

\usepackage[version=3]{mhchem} 

\usepackage{dcolumn}
\usepackage{bm}
\usepackage{todonotes}
\usepackage{tikz}
\usepackage{nicefrac}
\usepackage{array}
\usepackage{amssymb}
\usepackage{amsmath}
\usepackage{graphicx}
\usepackage{multirow}
\usepackage{color}
\usepackage{comment}
\usepackage[normalem]{ulem}
\usepackage{url}
\usepackage{mathrsfs}
\usepackage{subfigure}
\usepackage{float}
\usepackage{tabularx}

\usepackage{siunitx}

\newcommand{\rev}[1]{{\color{black}{#1}}}
\newcommand{\revi}[1]{{\color{black}{#1}}}

\newcommand*{\citen}[1]{%
  \begingroup
    \romannumeral-`\x 
    \setcitestyle{numbers}%
    \cite{#1}%
  \endgroup   
}

\author{Yair Litman}
\affiliation[FHI]{Theory Department, Fritz Haber Institute of the Max Planck Society, Faradayweg 4--6, 14195 Berlin, Germany}

\author{Jeremy O. Richardson}
\affiliation[ETH]
{Laboratory of Physical Chemistry, ETH Zurich, 8093 Zurich, Switzerland}

\author{Takashi Kumagai}
\affiliation[FHI2]{Physical Chemistry Department, Fritz Haber Institute of the Max Planck Society, Faradayweg 4--6, 14195 Berlin, Germany}

\author{Mariana Rossi}
\email{rossi@fhi-berlin.mpg.de}
\affiliation[FHII]{Theory Department, Fritz Haber Institute of the Max Planck Society, Faradayweg 4--6, 14195 Berlin, Germany}

\title{ Elucidating the \rev{Nuclear} Quantum Dynamics of Intramolecular Double Hydrogen Transfer in Porphycene}


\begin{document}
\begin{tocentry}




    \centering
    \includegraphics[width=0.95\textwidth]{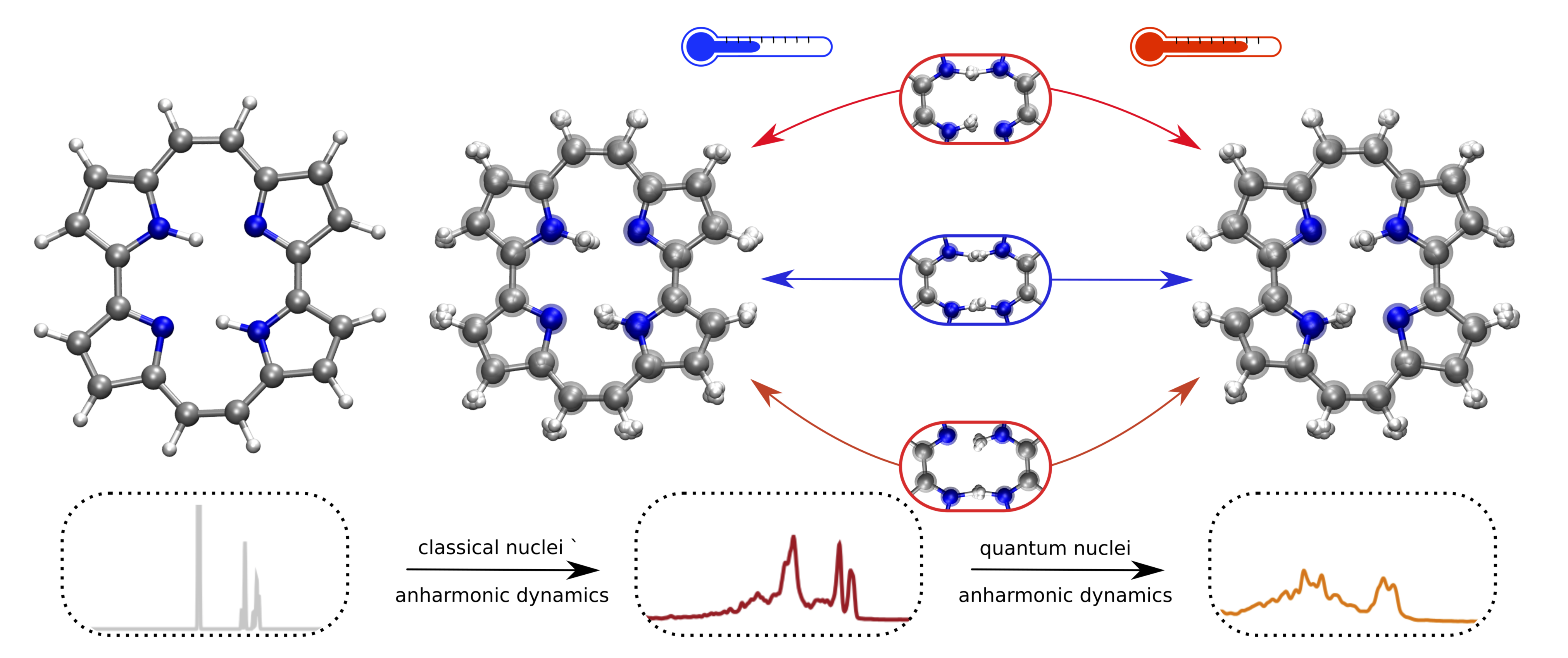}
\end{tocentry}

\begin{abstract}
\rev{We address the double hydrogen transfer (DHT) dynamics of the porphycene molecule: A complex paradigmatic system where the making and breaking of H-bonds in a highly anharmonic potential energy surface requires a quantum mechanical treatment not only of the electrons, but also of the nuclei. We combine density-functional theory calculations, employing hybrid functionals and van der Waals corrections, with recently proposed and optimized path-integral ring-polymer methods for the \revi{approximation} of \revi{quantum} vibrational spectra and reaction rates. Our full-dimensional ring-polymer instanton simulations show that below 100 K the concerted DHT tunneling pathway dominates, but between 100 K and 300 K there is a competition between concerted and stepwise pathways when nuclear quantum effects are included. We obtain ground-state reaction rates of $2.19 \times 10^{11} \mathrm{s}^{-1}$ at 150 K and $0.63 \times 10^{11} \mathrm{s}^{-1}$ at 100 K, in good agreement with experiment. We also reproduce the puzzling N-H stretching band of porphycene with \revi{very good} accuracy from thermostatted ring-polymer molecular dynamics simulations. The position and lineshape of this peak, centered at around 2600 cm$^{-1}$ and spanning 750 cm$^{-1}$, 
\revi{stems from a combination of very strong H-bonds, the coupling to low-frequency modes, and the access to {\it cis}-like isomeric conformations, which cannot be appropriately captured with classical-nuclei dynamics.}
These results verify the appropriateness of our general theoretical approach and provide a framework for a deeper physical understanding of hydrogen transfer dynamics in complex systems.}
\end{abstract}

\section{Introduction}

In \rev{hydrogen and proton transfer reactions} \cite{HYDROGEN_TRANSFER}, electronic polarization and anharmonicities of the potential energy surface (PES) cause a considerable coupling between different degrees of freedom in the system, leading to complex intra- and intermolecular vibrational energy transfer\cite{Warshel_1982_JPC}. In addition, due to the light mass of hydrogen, nuclear quantum effects (NQEs) such as tunneling and zero-point energy (ZPE) \rev{can} play a crucial role at remarkably high temperatures \cite{Tuckerman_Science_1997, Marx_1999_Nature, Klinman_AnnRevBio_2013, Meisner_Angew_2016, RossiMano2016},
and even heavy atoms (e.g., carbon or oxygen) can have an impact on \rev{hydrogen tunneling} \cite{Tuckerman_PRL_2001,Hinsen_JCP_1997}. 
\rev{An accurate and quantitative description requires an all-atom, all-electron quantum simulation of these reactions, and makes theoretical approaches based on empirical potentials and/or a dimensionality reduction of the PES, as well as a classical treatment of nuclear dynamics inadequate.}

Porphycene, the first synthesized structural isomer of free-base porphyrin \cite{Vogel_AngChem_1986}, 
has emerged as a unique example of \rev{intramolecular double hydrogen transfer (DHT)} in a multidimensional anharmonic PES, where strong hydrogen bonds are formed in the molecular cavity \cite{Fita_PCCP_2017,Waluk_ChemRev_2017}. 
\revi{The interest in this molecule has increased in the recent years, especially due to results obtained from scanning probe microscopy. 
Those have shown how different external stimuli, such as light \cite{Bockmann_NanoLett_2016}, heat \cite{Kumagai_PRL_2013}, and force\cite{Ladenthin_NAT_2016} can control the DHT, establishing this molecule as a prototype for the manufacture of rapid optical molecular switches and logic gates. }

\begin{figure}
    \centering
    \includegraphics[width=0.62\textwidth]{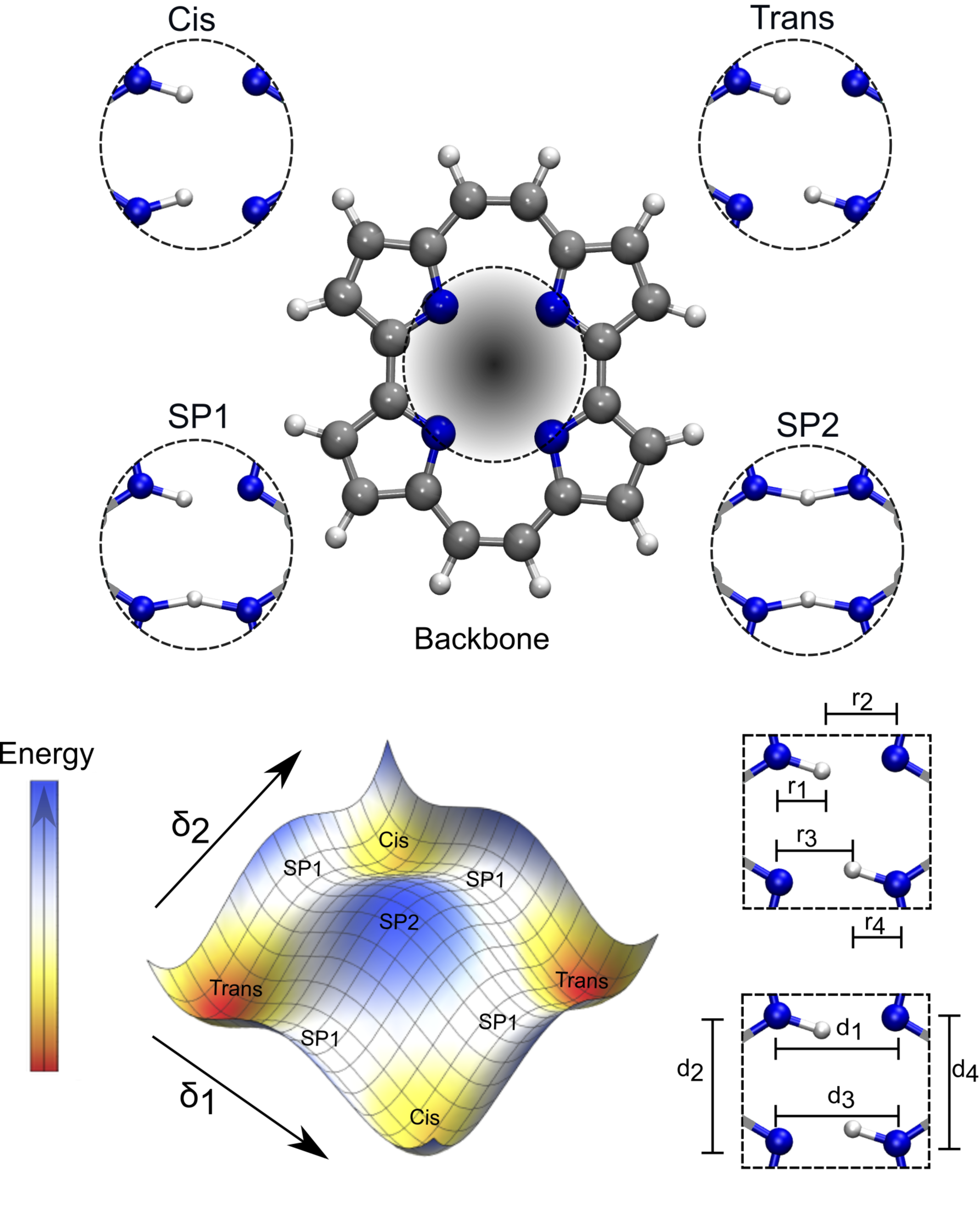}
    \caption{Characterization of the \textit{cis} and \textit{trans} isomers as well as a schematic 2D potential energy surface (PES) of the porphycene molecule. SP1 and SP2 stand for first order and second order saddle points and $r$ and $d$ are distances between atoms as marked in the figure. \revi{In the schematic 2D PES}, $\delta_1=r_1-r_2$ and $\delta_2=r_3-r_4$.}  
    \label{fig:Porphycene}
\end{figure}

The presence of \revi{strong} hydrogen bonds results in a low reaction barrier and a high DHT 
rate \rev{at room temperature} \rev{($k\approx10^{12}$ s$^{-1}$)} \cite{Fita_2009_CEJ} compared to other derivatives, e.g. porphine ($k\approx10^{4}$ s$^{-1}$) \cite{Braun_JACS_1994}. 
The DHT reaction, which in this case connects the two degenerate {\it trans} tautomeric states (Fig. 1), can occur either through a stepwise or concerted mechanism. 
In the former case, the hydrogen atoms are transferred sequentially \revi{through an} intermediate {\it cis} tautomeric state and \revi{the reaction pathway involves} a first-order saddle point (SP1 in Fig. \ref{fig:Porphycene}). 
\revi{In the latter case, two hydrogen atoms are transferred in a correlated fashion without passing by a stable intermediate. This transfer can occur simultaneously through a second-order saddle point (SP2 in Fig. \ref{fig:Porphycene}), which is usually referred to as the synchronized mechanism\cite{Dewar_JACS_1984}}. 
\rev{It is also well established that the tautomerization reaction in the porphycene molecule is a multidimensional process. The strong vibrational level dependence of the tunnelling splittings \cite{Mengesha_2013_JCP,Vdovin_2009_CPC} and the modification of the tautomerization rate upon deuteration of the peripheral hydrogens \cite{Mengesha_JPCB_2015} are two manifestation of the multidimensional nature of the tunneling process.}
\rev{At cryogenic temperatures in the gas-phase (or in a helium droplet), the vibrational ground state level exhibits a tunneling splitting of 4.4 cm$^{-1}$ \cite{Vdovin_2009_CPC,Mengesha_2013_JCP,Sepiol_1998_CPL}, 
showing that the {\it trans-trans} tautomerization takes place via coherent tunneling through the concerted mechanism  \cite{Smedarchina_JCP_2007,Smedarchina_JCP_2014}. Indeed, excellent agreement with the experimental values have been obtained considering only this pathway\cite{Zahra2014}.}
However, \revi{vibrational couplings to the environment or, at higher temperatures, couplings to the intramolecular vibrational modes} \rev{lead to decoherence of the wave-function such that the quantum state decays exponentially and the reaction dynamics can be described} as a rate process \cite{Ciacka_2016_JPCL}. In these cases, which are more relevant for biological or technological applications,
it is not straightforward to clarify which mechanism governs the DHT reaction, how temperature affects the reaction rates, and the character of the vibrational modes that play a role.

Vibrational spectroscopy is a powerful tool that can shed light on nuclear dynamics and the PES. \rev{The hydrogen-stretching} vibrational band in this system \rev{serves as a fingerprint of hydrogen bonding and can} 
exhibit extremely complex spectral features, such as a  broadening and the emergence of satellite peaks.
In porphycene, the characterization of the N-H stretching mode ($\nu_{\text{N-H}}$) related to the  {\it trans-trans} tautomerization has been particularly controversial, because of its apparent absence in the experimental Raman and IR spectra, although theoretical calculations in the harmonic approximation predict a very intense band \cite{Gawinkowski_2012_PCCP}. The broadening of $\nu_{\text{N-H}}$ in porphycene resulting from the intermode coupling has been examined by a classical nuclear dynamics approach \cite{Gawinkowski_2012_PCCP}, but an agreement between simulations and experiment could not be fully achieved. This suggests that a strongly anharmonic PES, intermode couplings, and NQEs \rev{all} play a decisive role in the static and dynamic properties of the inner hydrogen atoms of porphycene.  

The situation outlined above has hampered the elucidation of the hydrogen-bonding structure and tautomerization dynamics in porphycene for a long time. \rev{\revi{Approximations to the} quantum dynamics have been \revi{reported} but they relied on a dimensionality-reduction of the system 
\rev{and thus neglected vibrational coupling to several modes}
\cite{SmedarchinaKuhn2007,ShiblKuhn2007,Smedarchina_JCP_2014,McKenzie_JCP_2014}. 
Classical molecular dynamics simulations have been performed which included the full dimensionality of the system, but neglected the quantum nature of nature of the nuclei \cite{Walewski_JPCA_2010,Gawinkowski_2012_PCCP}.
Imaginary-time path-integral simulations based on a semiempirical PES have been carried out to compute quantum statistical properties,
but did not use methodology that could give access to real-time dynamics of the system (e.g. rates or vibrational spectra)
\cite{Yoshikawa_CP_2012,Yoshikawa_CPL_2010}. To date, an atomistic full-dimensional \revi{study} of the quantum dynamics in this system has not been presented.}

Here we demonstrate that all-atom and all-electron quantum simulations of \rev{porphycene} can fully reveal the equilibrium and dynamical properties of this molecule in the gas-phase and explain a variety of experimental observations. \rev{Employing newly implemented and optimized ring-polymer instanton theory\cite{i-pi2} and the recently proposed ring polymer molecular dynamics coupled to colored noise thermostats\cite{Rossi_JCP_2018},} 
we assess in detail the open questions regarding the hydrogen bonding geometry, the hydrogen transfer dynamics, and the nature of the $\nu_{\text{N-H}}$ band observed in the IR spectrum \rev{using a first-principles PES and without resorting to any dimensionality-reduction}. 
We show that the hydrogen transfer mechanism in porphycene follows non-intuitive pathways at certain temperatures, which underlines the significance of NQEs and intramolecular vibrational coupling.

\section{Results and discussion}

\begin{figure}
    \centering
    \includegraphics[width=0.8\textwidth]{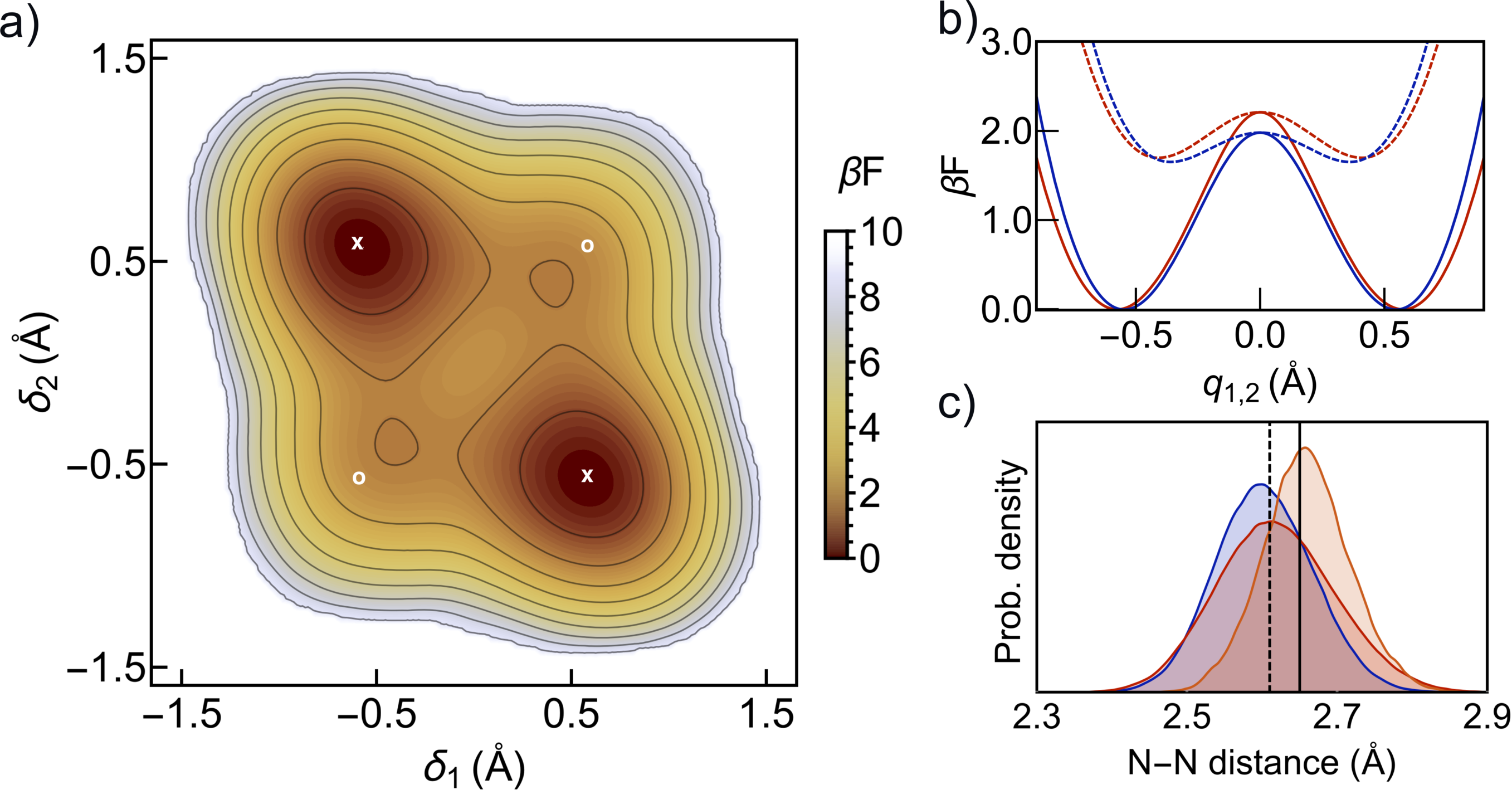}
    \caption{(a) Effective free energy profile of the porphycene molecule obtained from PIMD simulations at 290K, projected on the $\delta_1$ and $\delta_2$ coordinates.  The contour lines start at 0.75 $k_\mathrm{B}T$ and are separated by 1 $k_\mathrm{B}T$. The x and o symbols mark the position of the \textit{trans} and \textit{cis} conformers, respectively, when optimized at the potential energy surface. \rev{We define ``{\it cis}-like" conformations as the ones lying in the vicinity of ($\delta_1,\delta_2$) equal to (+0.4,+0.4) or (-0.4,-0.4)}
    (b) Effective free energy projections along  $q_1$ ($\delta_1 = -\delta_2$) (solid lines) and $q_2$  ($\delta_1 = \delta_2$) (dashed lines) directions. Red and blue lines correspond to PIMD simulations at 290K and 100K respectively.  (c) Nitrogen-nitrogen distance probability density obtained from PIMD simulations at 100K (blue), PIMD simulations at 290K (red) and MD simulation at 290K (orange). Vertical lines indicate the minimum energy geometry values for the {\it trans} isomer (solid) and  {\it cis} isomer (dashed).}
    \label{fig:FreeEnergy}
\end{figure}

We first analyze the equilibrium properties of the porphycene molecule obtained from density-functional theory (DFT-B3LYP+vdW) calculations and \textit{ab initio} path-integral molecular dynamics (PIMD) simulations (see Methods). 
In Fig.\ \ref{fig:FreeEnergy}a we show the free energy profile projected on the two hydrogen transfer coordinates $\delta_1$ and $\delta_2$ defined in
Fig.\  \ref{fig:Porphycene} for the PIMD simulation at 290 K\@. 
As expected, the {\it trans} isomer is the most energetically stable and the free energy barrier for hydrogen transfer is around  $2 k_\mathrm{B}T$, in agreement with previous studies where NQEs were also included \cite{Yoshikawa_CPL_2010, Yoshikawa_CP_2012}. For comparison, in the classical nuclei \textit{ab initio} molecular dynamics (MD) simulation at the same temperature, the hydrogen transfer reaction is a much rarer event with an effective barrier above $4 k_\mathrm{B}T$, as shown in the SI. 

The stationary points of the free-energy surface can be assigned to the {\it cis}, {\it trans}, first order (SP1) and second order saddle point (SP2) states of the molecule. 
\revi{While the exact boundary value ($b$) to distinguish different states is arbitrary due to the non-negligible density at all values connecting both states within the PIMD simulations, the {\it cis/trans} population ratio is robust for reasonable values $0.1<b<0.3$, as shown in Fig. S5 in SI.}
We calculate the {\it trans} population to be larger than the {\it cis} population by a factor 7. \rev{Note that these 2D free-energy surfaces are provided as a guide to understand this system. All our calculations are carried out in the full $3N$-dimensional space.}

The 1D PIMD free-energy projections along the $q_1$ and $q_2$ coordinates (defined along the lines $\delta_1=-\delta_2$ and $\delta_1=\delta_2$, respectively) \rev{are} shown in 
Fig.\ \ref{fig:FreeEnergy}b at two temperatures, namely 290 K and 100 K\@. 
These projections exhibit two interesting features. First, the effective energy difference between the wells, which within a first approximation would determine the {\it cis} and {\it trans} populations, does not change between 290 and 100 K. This means that in this temperature range only the lowest vibrational modes involving the N-H groups are populated and consequently the relative population is mostly governed by ZPE\@. The wells along $q_2$ ({\it cis}), however, are considerably softened towards the barrier when the temperature is reduced while the wells along $q_1$ ({\it trans}) are not. This effectively increases the probability of {\it cis} conformations at lower temperatures. Second, lowering the temperature has the surprising effect of decreasing the effective \rev{\textit{trans-trans}} free energy barrier by 10\% \rev{and the \textit{cis-cis} by 35\%.} This effect can be traced back to fluctuations in the cage dimensions. In Fig. \ref{fig:FreeEnergy}c we show the probability density of the N-N distances, $d_1$ and $d_3$, defined in Fig.\ \ref{fig:Porphycene}. 
The quantum distribution at 290 K is broader and shifted to higher values in comparison to the one at 100 K\@. The cage expansion at higher temperatures leads to an increase in the effective potential energy barrier for the hydrogen transfer reaction and is directly translated to the free energy profile. 
The smaller average N-N distances at lower temperatures correlates with an increased amount of $cis$-like conformations with respect to $trans$, and is consistent with the strengthening of the hydrogen bonds observed from previous calculations on a reduced PES and NMR experiments \cite{ShiblKuhn2007}. 
\rev{In our PIMD simulations we also observe that the average absolute value of the dipole projected along the H-bonded N atoms is 0.5 Debye which is in contrast to the absent dipole moment of the {\it trans} minimum-energy isomer. At the same time, it is also lower than the corresponding dipole of the {\it cis} minimum-energy isomer (1.2 Debye). This observation is consistent with the observed N-N distance distribution shown in Fig. \ref{fig:FreeEnergy}c, which is centered closer to the N-N distance of the optimal {\it cis} isomer when considering quantum nuclei.}
In the classical-nuclei simulations, the \textit{cis}-like geometries are not visited at all, and the cage is much larger. 

At this point, it is worth noting that a reduction of the effective free energy barrier along the chosen reaction coordinate does not necessarily result in an \emph{increase} on the hydrogen transfer rate. Within transition state theory
(TST), a probabilistic term that depends on the effective barrier height is multiplied by a dynamical correction factor (also called transmission coefficient). It has been shown that the dynamical factor can have a much stronger temperature dependence behaviour than the probabilistic one \cite{Craig_JCP_2005_2}. Consequently, rates and barrier heights are not always trivially correlated and the temperature dependence of the rate obtained based exclusively on free energy barriers should be \rev{treated with caution}.
In fact, for this system it is experimentally observed that the rates \emph{decrease} with decreasing temperatures \cite{Ciacka_2016_JPCL} and this is also what we find in our \rev{full-dimensional} calculations shown below. 

We now examine the quantum dynamical aspects of the hydrogen transfer reactions, which we \revi{approximate} with ring-polymer instanton theory \cite{Miller_JCP_1975,Jor_perspective_2018} and thermostatted ring-polymer molecular dynamics \cite{Rossi_JCP_2014}. We \rev{located} two instanton pathways, namely, the one related to the {\it trans-trans} concerted pathway and the one related to the {\it trans-cis(-trans)} stepwise pathway.
\rev{Each instanton is a uniquely-defined path which best describes a tunneling mechanism in porphycene.}
With these pathways, we can compute the thermal rate as $k= 2k_\text{step}+k_\text{conc}$ \cite{SmedarchinaKuhn2007},
where $k_\text{step}$ and $k_\text{conc}$ are the rate constants corresponding to the concerted and stepwise mechanisms, respectively. The factor 2 takes into account the two possible \textit{cis} intermediates. The ratios $R_s=2k_\text{step}/k$ and $R_c=k_\text{conc}/k$ show the relative contribution of each pathway to the total observed double-hydrogen transfer rate.  
\rev{The fact that we find two stable instanton pathways means that there are at least two independent pathways for the DHT. Even though we cannot guarantee that all possible DHT tunneling pathways are being considered, the symmetry of porphycene indicates that these should be the dominating (or the only) ones.}
Our best estimate for the DHT rates are presented in Table \ref{tab:rates}. These values are in good agreement with the experimental values measured in solution. 
\rev{Direct comparison between our gas-phase calculation and these results in certain condensed-phase environments is supported by the observation of negligible solvent dependency in experiments \cite{Ciacka_2015_JPCB,Gil_JACS_2010}.}
\revi{In other types of environment such as surfaces \cite{Koch_JACS_2017,Kumagai_JCP_2018} or polymer matrices \cite{Piatkowski_2018_JPCL}, the rate depends more strongly on the environment. Larger porphycene derivatives show rate dependency in both isotropic\cite{Gil_JACS_2010} and non-isotropic environments \cite{Piwonski_JPCL_2013}.}

The tunneling enhancement factor, also reported in Table \ref{tab:rates}, is defined 
as $\kappa_{\text{tun}}=k_\text{inst}/k_{\text{TST}}$, where $k \equiv k_{\text{inst}}$ is the rate obtained with the instanton method and $k_{\text{TST}}$ is the rate obtained with the Eyring \revi{transition state theory} (TST)\@. 
\rev{As expected,} tunnelling effects thus increase the rates by almost 2 and 3 orders of magnitude at 150 K and 100 K, respectively.
More interestingly, at $T$=150 K the concerted and stepwise mechanisms have a comparable contribution to the total rate, while at $T$=100 K the concerted path accounts for 95\% of the rate. Note that if we neglect NQEs, the scenario is completely different and at both temperatures the stepwise \revi{mechanism} fully dominates. 

\begin{table}
\centering
	\begin{tabular}{|c|c|c|c|c|c|c|}
	\hline
	& $T$ (K)& $k$  $(10^{11} \mathrm{s}^{-1})$ & $k_{\text{exp}}$ $(10^{11} \mathrm{s}^{-1})$ & $R_c$ & $\kappa_{\text{tun}}$ & KIE\\
	\hline
    HH & 150 & 2.190 & 1.62$\pm$0.04  & 0.63 & 70 &  \\
    HH & 100 & 0.630 & 1.09$\pm$0.08  & 0.95 & 1135 &  \\
    DD & 100 & 0.0025 & 0.03$\pm$0.02 & 0.59 & 471   & 250\\
    \hline
	\end{tabular}
	\caption{Our calculated hydrogen transfer rates $k$, compared to experimental $k_{\text{exp}}$ data from Ref. \cite{Ciacka_2016_JPCL}, as well as the tunneling enhancement factor $\kappa_{\text{tun}}$, the contribution from the concerted path to the rate $R_c=k_\text{conc}/k$, and the calculated kinetic isotopic effect KIE=$k^{\text{HH}}/k^{\text{DD}}$}\label{tab:rates} 
\end{table}

\begin{figure}
    \includegraphics[width=\textwidth]{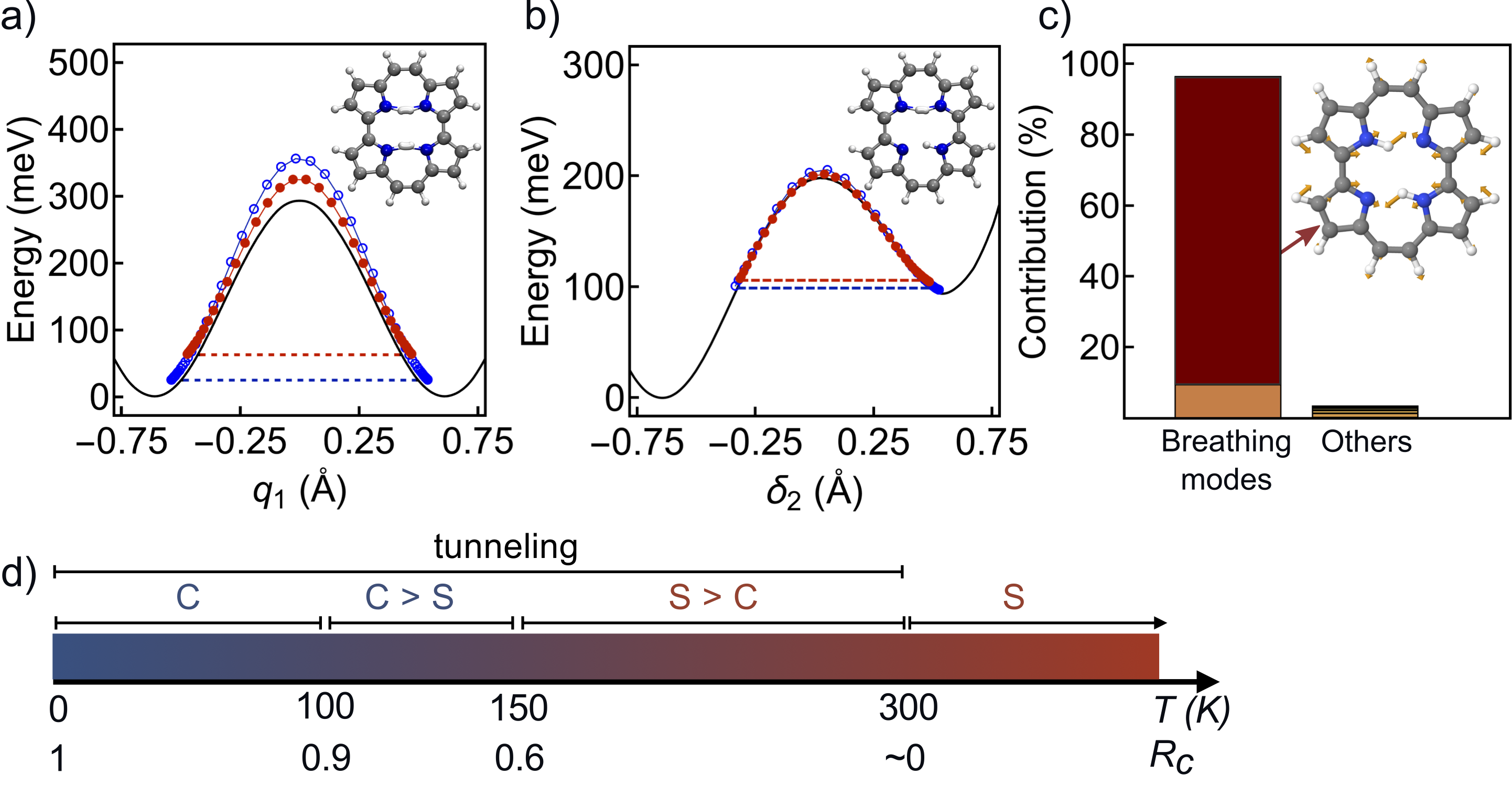}
    \caption{(a) Potential energies of different instanton imaginary-time slices along the \textit{trans-trans} path at 100 K (open blue circles) and 150 K (filled red circles). The black energy profile marks the ground state potential energy surface and the dashed lines mark the minimum energy \revi{of the intanton slices}. (b) Same as in a for the \textit{trans-cis} path. (c) Decomposition of the instanton turning point displacement of the \textit{trans-trans} path into the mimimum energy \textit{trans} geometry normal modes. The inset shows the breathing mode at 197 cm$^{-1}$  (2A$_g$) with largest contribution. (d) Competition of different transfer mechanisms at different temperatures: C denotes the concerted mechanism and S the stepwise mechanism.
    $R_c$ is the contribution from the concerted path to the rate as defined in the caption of Table \ref{tab:rates}.
    }  
    \label{fig:Mechanism}
\end{figure}

This result can be understood in terms of the mechanisms predicted by the instanton pathways.
In Fig.\ \ref{fig:Mechanism}a and b we show the minimum energy path (MEP) for the concerted {\it trans-trans} path and the stepwise {\it trans-cis} path, respectively, at 150 and 100 K\@ (\revi{black solid curves}). 
The concerted path has a higher barrier than the stepwise one.  Analyzing the tunneling paths themselves, we note that the stepwise path requires that the initial \textit{trans} geometry \rev{is thermally excited to} an energy comparable to the intermediate \textit{cis} geometry, which is around 100 meV higher than the minimum, in order for tunneling to take place. From that point, the tunnelling path follows quite closely the MEP and has essentially no temperature dependence between 100 and 150 K\@. In contrast, \rev{the concerted path enables tunneling at an energy close to the bottom of the well. The energy at which the system tunnels in this mechanism is temperature dependent
and is defined by instanton theory to give the optimal balance of the probability of barrier penetration and the probability of thermal excitation}.\cite{Jor_perspective_2018}.
At 100 K the tunnelling path starts at 25 meV above the minimum and reaches points above 350 meV\@. At 150 K, since more thermal energy is available, the path starts at 64 meV but does not reach such high values, keeping closer to the MEP. 

At low temperature, only the concerted mechanism is possible since there is not enough thermal energy available to reach the high energy conformation at which the stepwise path takes place. The concerted mechanism presents a larger effective path length because two particles need to tunnel and consequently the microcanonical tunnelling probability is always smaller than for the stepwise case (See SI). For this reason, increasing the available thermal energy strongly favors the stepwise mechanism. Indeed, above the crossover temperature of this system \cite{Gillan1987}, $T_\mathrm{c}=\hbar\omega^{*}/2\pi k_\mathrm{B}\approx 300$ K, where 
\rev{in most cases the contribution to the rate from classical hopping over the barrier becomes more important than tunneling, the stepwise mechanism is dominant. Previous evidence of the stepwise mechanism at 300 K from classical-nuclei molecular dynamics simulations reported in Ref. \citen{Walewski_JPCA_2010} benefit from this fact, but those simulations would not be able to predict the intricate balance between the two mechanisms at lower temperatures.} Our calculated DHT mechanisms at different temperatures are \rev{summarized} schematically in Fig.\ \ref{fig:Mechanism}d. 

The competition of both mechanisms below $T_\mathrm{c}$ is compatible with some previous 2D-model studies \cite{Smedarchina_JCP_2007}. However, it goes against the common experimental 
interpretation that this reaction happens exclusively through the concerted path at any (low) temperature \cite{Mengesha_JPCB_2015,Gil_JACS_2007}. In particular, the lack of {\it cis} conformers observed in 
fluorescence anisotropy studies is treated as an indirect evidence for the concerted mechanism \cite{Ciacka_2015_JPCB,Fita_2009_CEJ}. We propose that this is not a conclusive observation. The 
energy difference between the two isomers results in a very small {\it cis} population and due to the low barrier height for the {\it cis-trans} reaction, any transient {\it cis} isomer will have a very short lifetime. Therefore, 
anisotropy measurements performed so far cannot rigorously confirm the mechanism in this particular scenario. Another possible experimental strategy to discern the DHT mechanism is to monitor kinetic isotope effects (KIE) upon substitution of two (DD) internal hydrogens by deuterium \cite{Braun_JACS_1994}. 
However, due to the lack of well defined $\nu_{\text{HH}}$ and $\nu_{\text{DD}}$ bands, it has been difficult to interpret and quantify the isotopic purity of any sample. 
\revi{The KIE is easily accessible from out calculations and}
we report in Table \ref{tab:rates} the DHT rate and the KIE for the DD \rev{isotopologue} of porphycene at 100 K.
KIEs from instanton theory benefit from error cancellations \cite{hexamerprism} that make them typically more reliable than direct rate calculations.
Nonetheless, our calculated KIE is 7 times higher than what was reported from experiments \cite{Ciacka_2016_JPCL}. \revi{This discrepancy would be consistent with an error of the theoretical barrier width with respect to experiment, especially for the stepwise mechanism. Indeed, we are correcting the B3LYP+vdW barrier height with data from CCSD(T), but we cannot correct the barrier width with the available data. Since the {\it trans-trans} pathway is symmetric and the {\it trans-cis} (stepwise) pathway is not, correcting only the barrier should work better for the former. The deuterium rate at 100 K involves a large fraction of the stepwise-mechanism contribution and thus is prone to a larger error.}
\rev{From our simulations, we find that} the DHT mechanism itself changes upon isotopic substitution, with deuteration strongly favoring the stepwise path as reflected by $R_c$. 
This situation is different for porphyrin, where there is no competition between the different pathways already for HH, such that it was possible to confirm the dominance of the stepwise mechanism from experiments \cite{Braun_JACS_1994}.

Our full dimensional description allows us to go one step further and understand mode-coupling in this DHT in more detail. \rev{We employ the normal-mode decomposition proposed in Ref.\ \citen{Zahra2014} to analyze mode-specific contributions to tunneling splittings in porphycene. We take} the \textit{trans} isomer normal modes to
decompose the displacement between the geometry from where the tunneling takes place (the instanton turning point) of the concerted path and the global minimum. 
Our decomposition, shown in Fig. \ref{fig:Mechanism}c, shows that more than \rev{87\%} of the displacement is given by the breathing mode of the molecular cage at around 200 cm$^{-1}$ (\revi{24 meV}). The coupling of the \rev{concerted} DHT to this low-energy mode implies a temperature-dependence for this reaction rate at low-temperatures that can look like an Arrhenius behavior, despite the massive contribution from tunneling. 
\rev{Therefore, we show that the concerted mechanism is related to the} temperature dependence of the rate at low temperatures measured in Refs. \citen{Ciacka_2016_JPCL} and \citen{Gil_JACS_2007}, which is consistent with a barrier of 22 meV and the previous suggestion that this vibrational mode contributes to the DHT\@.
\rev{A novel insight that we gain from our simulations is that we can unambiguously explain the origin of the activation barrier for the second thermally activated channel reported in Ref.\ \citen{Ciacka_2016_JPCL}. From Fig.\ \ref{fig:Mechanism}b, it is apparent that for the stepwise tunneling path to be activated, the {\it cis} conformer minimum-energy needs to be reached. Detailed balance thus imposes an effective activation energy for this mechanism equal to the {\it trans} to {\it cis} energy difference. Our best estimate for this value is 107 meV (CCSD(T) with MP2 ZPE corrections), which matches the experimentally inferred activation barrier of 108 $\pm$ 9 meV.}

\begin{figure}
    \centering
    \includegraphics[width=0.9\columnwidth]{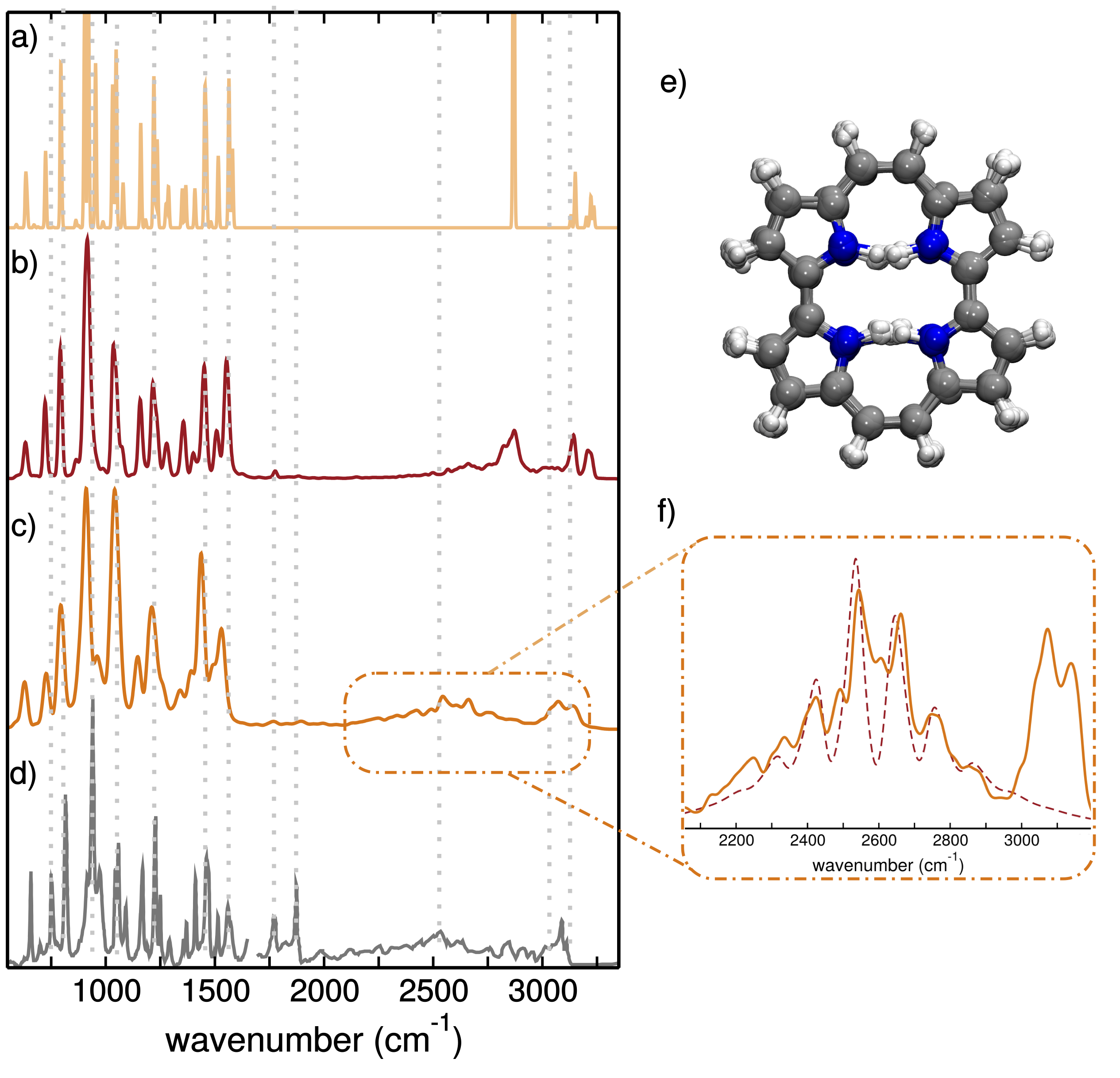}
    \caption{Porphycene IR spectra obtained through the (a) harmonic approximation, (b) Fourier transform of the dipole autocorrelation function from DFT-B3LYP+vdW trajectories with classical nuclei at 290 K, (c) Fourier transform of the dipole autocorrelation function from DFT-B3LYP+vdW trajectories with quantum nuclei at 290 K from TRPMD. (d) Experimental spectra as reported in Ref. 27. \revi{Below 1700 cm$^{-1}$ this spectrum was measured for porphycene in a CS$_2$ solution and above 1700 cm$^{-1}$ for porphycene in a KBr matrix.}
    The harmonic spectrum was artificially broadened with Gaussian functions for better visualization \revi{and all the theoretical spectra have a constant red shift of 40 cm$^{-1}$}. Dotted lines serve as guides to the eye. 
    (e) Representative snapshots of the trajectory with quantum nuclei where all the replicas are superimposed. (f) Zoom of the Fourier transform of the \revi{TRPMD IR spectrum} in the N-H stretching mode region (orange) together with the adiabatic model (red dashed line) described in the text. }
    \label{fig:IR-B3LYP}
\end{figure}

The signature of anharmonic intermode coupling and quantum contributions in the DHT is directly reflected in the vibrational fingerprints that appear in an IR spectrum. We compare in Fig.\ \ref{fig:IR-B3LYP} the IR spectra obtained with the quantum harmonic approximation (QH), the (anharmonic) time-correlation formalism with classical-nuclei dynamics (CL) at 290 K, and the same formalism with \revi{the approximation to} quantum-nuclei dynamics \revi{from TRPMD} (QA) at the same temperature (see Methods). We also present the experimental IR spectrum reported \revi{in Ref. \citen{Gawinkowski_2012_PCCP}. Below 1700 cm$^{-1}$ this spectrum was measured for porphycene in a CS$_2$ solution and above 1700 cm$^{-1}$ for porphycene in a KBr matrix.}

Our calculated QH and CL spectra are in good agreement to the theoretical simulations previously presented in Ref. \cite{Gawinkowski_2012_PCCP}. 
As discussed there, (classical-nuclei) anharmonicities play an important role for  $\nu_{\text{N-H}}$ at around 2900 cm$^{-1}$ (easily assignable from normal mode analysis) and coupling to low-frequency vibrations considerably broaden the peak.  
However, the band is still present and shows a different shape than the experimentally measured spectrum at 290 K from Ref. \citen{Gawinkowski_2012_PCCP}. 
With the inclusion of NQEs, this peak is red-shifted by 300 cm$^{-1}$ and broadened, producing a very weak signal, remarkably similar to the experimental line shape. We only observe a constant shift of \revi{$\approx$40 cm$^{-1}$} between our simulations and experiment, which likely originates from the B3LYP+vdW PES. We corroborated the NH-stretch character of this peak by comparing spectra from hydrogenated and deuterated molecules (see SI, Figure S9). The overtone peaks just below 2000 cm$^{-1}$ that were also observed in Ref. \citen{Gawinkowski_2012_PCCP} are present in our CL and QA simulations, albeit with a much lower intensity than the experimental bands. \revi{This discrepancy in intensities could stem from the fact that both CL and QA spectra cannot capture quantum coherence. It could also stem from a particular enhancement of these bands due to the KBr matrix present in the experiment or to the laser power profile. In order to resolve this issue, comparison to gas-phase experimental spectra at lower temperatures would be desirable, but such data is not available. Overall, the present} results show that NQEs are essential to explain quantitatively the softening of the $\nu_{\text{N-H}}$ band and resolve its apparent absence in the experimental spectrum \cite{Gawinkowski_2012_PCCP, Malsch_JPCA_1997}.

One can relate this lineshape also to the N-N distance distributions shown in Fig.\ \ref{fig:FreeEnergy}c. The shorter N-N distances in the path-integral simulations point to a strengthening of the H-bonds, which is related to the mode-softening (red-shift). Characterizing the strength of mode couplings and the fine-structure of this band, however, requires a deeper analysis, especially due to the fact that, as shown in Fig.\ \ref{fig:IR-B3LYP}e, the hydrogens are delocalized over the several energy wells. 

We follow the approach suggested in Ref.\ \citen{Blaise_ACP_2008}, that provides a way of calculating IR lineshapes for coupled low and high-frequency vibrational modes, that we detail in the SI. We used the adiabatic model including direct damping (relaxation of the high-frequency mode) and indirect damping (relaxation of the low-frequency mode which is coupled to the high-frequency one). The most relevant parameters of the model are the high ($\Omega$) and low ($\omega$) frequency values, as well as the strength of the coupling between them ($\alpha$). As shown in Fig.\ \ref{fig:IR-B3LYP}f, the agreement of the model with the TRPMD IR-spectrum is good, especially for the fine-structure of the mode. \revi{Since this model disregards hydrogen transfer, and since the DHT occurs in very long time-scales compared to the N-H vibration, we safely conclude that DHT does not contribute to the lineshape of this peak, similar to recent reports regarding the formic-acid dimer \cite{Qu_JCP_2018}}. The parameters for the curve shown in Fig.\ \ref{fig:IR-B3LYP}f are reported in the SI. We can obtain good agreement for $2680<\omega<2750$ cm$^{-1}$ and $100<\Omega<115$  cm$^{-1}$, with corresponding $\alpha$ ranging from 0.70 to 0.85. The exact value of $\Omega$ most likely reflects an effective value that averages the coupling to several low-frequency cage-vibration modes. The coupling parameter $\alpha$ is more than one order of magnitude larger than a simple evaluation in the strong anharmonic coupling theory \cite{Blaise_BOOK} based on harmonic modes, reflecting the enhancement of coupling due to anharmonic NQE. Notably, the $\omega$ values are very illuminating. They lie between the \textit{trans} (2907 cm$^{-1}$) and \textit{cis} (2680 cm$^{-1}$) harmonic frequency values. \revi{This shows the importance of visiting {\it cis}-like structures, which is only possible due to tunnelling and ZPE contributions.}

\section{Conclusions}

In this paper we have presented a combination of techniques joining high level density-functional theory simulations and \rev{state of the art} path-integral based approximations to nuclear quantum dynamics that provides a deep physical understanding of hydrogen transfer reactions. We validated the predictive power of our approach through comparison with experimental data.

In particular, we have shown that in an accurate potential energy surface, NQEs involved in hydrogen transfer events can {\it qualitatively} change dynamical properties \rev{of the porphycene molecule. Our calculated \rev{full-dimensional} ring-polymer instanton rate of DHT is in good agreement with experiment at different temperatures}. From our simulations, we are able to \rev{quantitatively obtain the} relative contribution from the stepwise and the concerted DHT mechanisms at any temperature. \rev{We conclude that both tunneling pathways have a} similar contribution \rev{to the rate} at 150 K\@ and that \rev{only below 100 K the concerted pathway becomes dominant. We corroborate that the skeletal low-frequency vibration that couples to the DHT coordinate leads to an Arrhenius-like temperature dependence for the rate, even when the DHT is dominated by tunneling. Additionally, we show that the contribution of the stepwise pathway down to low temperatures explains the second thermally activated reaction channel reported in experiments.} 

\revi{From the path-integral simulations}, we have shown how NQEs can strengthen N-H$\cdots$N bonds, modify the N-N distances, and shift peaks up to 300 cm$^{-1}$. The coupling and delocalization of the hydrogens within the cage explains the apparent absence of the N-H stretch band in the IR spectrum of this molecule, which had long puzzled researchers. \rev{The IR spectrum calculated from thermostatted ring-polymer molecular dynamics yields an exquisite agreement with the experimental lineshape of $\nu_{\text{N-H}}$, even for the fine-structure of the peak. We could reproduce this lineshape with a simple model of coupled quantum harmonic oscillators, which allowed us to confirm that the coupling of this mode to low frequency vibrations is enhanced when NQEs are included.}

\rev{A shortcoming of increasing the complexity is that a detailed quantitative understanding of vibrational mode-coupling can be compromised. \revi{Techniques like vibrational configuration interaction can, nevertheless, provide quantitative numbers even for rather complex molecules \cite{Chen_FarDis_2018}. Techniques based on dynamics, however, do not give direct access to coupling coefficients without the application of other models \cite{Geral_2011_JCTC}}. The \revi{adiabatic} model we employed in this work could give a qualitative explanation of the coupling and fine-structure of the NH-stretching mode in the IR spectrum. Mode-specific experimental and theoretical techniques, like time-resolved 2D-spectroscopy, would be necessary for a quantitative characterization of mode coupling at finite temperatures in high dimensional systems. Nevertheless, the approaches presented here pave the way to 
address important problems in biology, related to  enzymatic reactions through low-barrier H-bonds \cite{Cleland1998}, unusual fingerprints of N-H$\cdots$N bonds in proteins \cite{Adhikary2014}, as well as to model functional materials that take advantage of hydrogen transfer events \cite{Ueda_JACS_2014, Sunairi_JPCC_2018,Horiuchi_Nat_2010}.}

\section{Methods \label{sec:methods}}

We performed CCSD(T) calculations extrapolated to the infinite basis-set limit to obtain benchmarks for the energetics of local minima and saddle points of porphycene (see more details in SI). We then compared less expensive exchange-correlation (xc) functionals within density-functional theory (DFT) to these benchmarks, and concluded that the B3LYP functional \cite{B3LYP4} including pairwise van der Waals corrections (vdW) \cite{TS} presented the best agreement with our reference for both energetic and geometrical properties. We thus chose this functional to perform all further electronic-structure calculations in this work. The Orca package \cite{ORCA} was used to perform CCSD(T) simulations and the FHI-aims \cite{FHI-AIMS} package for DFT simulations.

The nuclear degrees of freedom were sampled according to the B3LYP+vdW PES with MD and PIMD through the i-PI program \cite{i-pi,i-pi2} in connection with FHI-aims. The PIMD simulations  were performed coupled to the colored noise PIGLET thermostat \cite{PIGLET}. We were therefore able to use 6 and 12 beads for 290 K and 100 K simulations, respectively.

Thermal rates were calculated with instanton rate theory \cite{Miller_JCP_1975,Jor_perspective_2018}.
The instanton path represents \revi{an approximation to} the optimal tunneling pathway at a given temperature and can be determined for the full high-dimensional problem. We used the ring-polymer approach \cite{Andersson2009Hmethane,Richardson_JCP_2009,Jor_review_2018,Rommel_JCTC_2011} where the path is discretized and represented by $P$ beads in a harmonic ring polymer \cite{Chandler_JCP_1981}, through our recent implementation in the i-PI code \cite{i-pi2}. 
We include all the 108 porphycene degrees of freedom in the path optimization. We have achieved convergence with 192 replicas at both 100 and 150 K (see SI). \rev{This corresponds to a simulation involving, effectively, 7296 atoms with first-principles and on-the-fly force evaluations.} We included a correction factor to the instanton action detailed in the SI, that linearly scales the action to match the CCSD(T) reference barrier, similar to Ref.\ \citen{Meisner:2018}.

Restarting an instanton calculation with higher number of replicas can be computationally expensive due to the necessity of 
calculating all Hessians for the new amount of replicas. In order to overcome this bottleneck, we have derived and implemented in i-PI a ring-polymer expansion of the Hessian, given by $\mathcal{H}^{(k)}_{jm} = \sum_{s=1}^{P}  H^{(s)}_{jm} T(P',P)_{ks}$, where $P<P'$ and $H^{(s)}_{jm}$ is the $j m$ matrix element of the Hessian corresponding to the \textit{s-th} original replica and $\mathcal{H}^{(k)}_{jm}$ is the $j m$ matrix element of the Hessian corresponding to the \textit{k-th} expanded replica. The $P' \times P $ matrix $T(P',P)$ is the same transformation matrix used in other contraction/expansion approaches and given in Ref.\citen{Markland_2008_JCP}. \rev{We provide the derivation of this expression in the SI}. This procedure completely removes the necessity of calculating \textit{any} new Hessians during the optimization, after their calculation for the first ring-polymer geometry with a small number of beads (replicas).

The IR spectra were computed from the Fourier transform of the dipole autocorrelation function. 
The classical correlation function was computed from 4 different 10 ps NVE trajectories and \revi{an approximation to the} quantum correlation function was computed using thermostatted ring polymer molecular dynamics \cite{Rossi_JCP_2014} coupled to generalized Langevin equation (TRPMD+GLE) thermostats, \revi{as described in Ref.\citen{Rossi_JCP_2018}}. In this case, we ran 7 different 10 ps trajectories using 16 beads (see convergence tests in the SI) and computed the dipole for all beads. In all cases, the starting configurations were taken from uncorrelated thermalized structures and we used a 0.5 fs time step for the integration of the equations of motion.

\begin{acknowledgement}
MR and YL acknowledge financial support from the Max Planck Society.
JOR's research is financially supported by the Swiss National Science Foundation (Project No.\ 175696).
\end{acknowledgement}

%
%

\bibliography{biblio}

\end{document}